\documentclass[slac_one]{revtex4}
\usepackage{graphicx}
\usepackage{fancyhdr}
\usepackage{bm}
\newcommand{\bnabla}{\bm{\nabla}}

\begin{document}
\arraycolsep=2pt
\title{How Does Quantum Vacuum  Energy Accelerate?} 

%

\author{K. A. Milton, P. Parashar, J. Wagner}
\affiliation{Oklahoma Center for High Energy Physics, and
H. L. Dodge Department of Physics and Astronomy, University of Oklahoma,
Norman, OK 73019 USA}
\author{K. V. Shajesh}
\affiliation{St.~Edwards School, Vero Beach, FL 32963 USA}
\author{A. Romeo}
\affiliation{Societat Catalana de F\'isica, Laboratori 
de F\'isica Matem\`atica (SCF-IEC), 08028 Barcelona, 
Catalonia, Spain}

\author{S. Fulling}
\affiliation{Departments of Mathematics and Physics, Texas A\&M University,
College Station, TX 77843 USA}

\begin{abstract}
We show that Casimir energy for a configuration of parallel plates
 gravitates according to the equivalence
principle both for the finite and divergent parts.  This shows that
the latter can be absorbed by a process of renormalization.
\end{abstract}

\maketitle

\thispagestyle{fancy}


\section{INTRODUCTION} 

The subject of Quantum Vacuum Energy (the Casimir effect) dates from the
same year as the discovery of renormalized quantum electrodynamics, 1948
\cite{casimir}.
It puts the lie to the naive presumption that zero-point energy is not
observable.
On the other hand, because of the severe divergence structure
of the theory, controversy has surrounded it from the beginning.
Sharp boundaries give rise to divergences in the local energy density near
the surface, which may make it impossible to extract meaningful self-energies
of single objects, such as the perfectly conducting sphere considered by
Boyer \cite{boyer}. 
These objections have recently been most forcefully presented by 
Graham, Jaffe, et al.~\cite{Graham:2003ib} and by Barton \cite{barton}, 
but they date back to 
Deutsch and Candelas \cite{Deutsch:1978sc,Candelas:1981qw}. 
In fact,
it now appears that these surface divergences can be dealt with successfully
in a process of renormalization,
and that finite self-energies in the sense
of Boyer may be extracted.

Gravity couples to the local energy-momentum tensor, and such
surface divergences promise serious difficulties. How does the 
completely finite Casimir interaction 
energy of a pair of parallel conducting plates, as well as the divergent
self-energies of non-ideal plates,
couple to gravity?

For a beginning of the renormalization of Einstein's equations
resulting from singular Casimir surface energy densities see 
Ref.~\cite{estrada}.

\section{GRAVITATIONAL ACCELERATION OF PARALLEL PLATE CASIMIR APPARATUS}

Brown and Maclay \cite{bm}
 showed that, for parallel perfectly
conducting plates separated by a distance $a$ in the $z$-direction, 
the electromagnetic stress
tensor acquires the vacuum expectation value between the plates
\begin{equation}
\langle T^{\mu\nu}\rangle
=\frac{\mathcal{E}_c}a \mbox{diag} (1, -1, -1, 3),
\quad \mathcal{E}_c=-\frac{\pi^2}{720 a^3}\hbar c.\label{t}
\end{equation}
Outside the plates the value of $\langle T^{\mu\nu}\rangle=0$.  

\subsection{Variational Principle}

Now we turn to the question of the gravitational interaction of
this Casimir apparatus.
It seems this question can be most simply addressed through
use of the gravitational definition of the energy-momentum tensor,
\begin{equation}
\delta W_m\equiv -\frac12\int(dx) \sqrt{-g}\,\delta g^{\mu\nu}T_{\mu\nu}
=\frac12\int(dx) \sqrt{-g}\,\delta g_{\mu\nu}T^{\mu\nu}.\label{var}
\end{equation}
For a weak field,
$g_{\mu\nu}=\eta_{\mu\nu}+2h_{\mu\nu}$,
(Schwinger's definition of $h_{\mu\nu}$).
Since the 
gravitational field is weak, we can ignore $\sqrt{-g}$.
The change in the
gravitational energy, for a static situation, is therefore given by
($\delta W=-\int dt\,\delta E$)
\begin{equation}
\delta E_g=-\int (d\mathbf{x}) \delta h_{\mu\nu}T^{\mu\nu}.\label{ge}
\end{equation}
If we replace $T^{\mu\nu}$ by $\langle T^{\mu\nu}\rangle$ in (\ref{t}), we
obtain an expression which is not gauge invariant, that is, which is dependent
on the coordinate system chosen, because $\partial_\nu \langle T^{\mu\nu}
\rangle\ne0$ on the plates.  To obtain a physically meaningful result,
we must adopt an interial coordinate system, which corresponds to the
use of the Fermi metric:
\begin{equation}
h_{00}=-gz,\quad h_{0i}=h_{ij}=0.
\end{equation}
Here, the  Cartesian coordinate
system attached to the earth is $(x, y, z)$, where
$z$ is the direction of $-\mathbf{g}$.  The Cartesian coordinates
associated with the Casimir apparatus (plate separation $a$,
length $L$) are $(\zeta,\eta,\chi)$, where $\zeta$
is normal to the plates, and $\eta$ and $\chi$ are parallel to the plates.
The angle between $\zeta$ and $z$ is arbitrary.

Now let the apparatus be rigidly displaced by an amount $\delta z$, a
constant.  Then we see that the gravitational force per area $A=L^2$ 
on the apparatus is independent of orientation:
\begin{equation}
\mathcal{F}=\frac{F}A=-\frac{\delta E_g}{A\delta z}=-g\int d\zeta\, T^{00}
=-g\mathcal{E}_c,
\label{f0}
\end{equation}
a small upward push.  This is exactly what the equivalence principle
would predict.  Thus the total energy of the Casimir apparatus,
consisting of the mass $M$ of the plates plus the Casimir energy,
undergoes the same acceleration,
\begin{equation}
\mathcal{F}_{\rm total}=-g(\mathcal{M}+\mathcal{E}_c),
\end{equation}
where the mass per area of the plates is $\mathcal{M}=M/A$.
For more details, see Ref.~\cite{Fulling:2007xa}.

\section{HYPERBOLIC MOTION}

Relativistically, uniform acceleration $1/\xi$
is described by hyperbolic motion
\begin{equation}
t=\xi\sinh\tau,\quad z=\xi\cosh\tau,
\end{equation}
which corresponds to the metric
\begin{equation}
dt^2-dz^2-dx_\perp^2=\xi^2d \tau^2-d\xi^2-dx_\perp^2.
\end{equation}
The d'Alembertian operator has in these coordinates a cylindrical form
\begin{equation}
-\left(\frac\partial{\partial t}\right)^2
+\left(\frac\partial{\partial z}\right)^2
=-\frac1{\xi^2}\left(\frac\partial{\partial\tau}\right)^2+\frac1\xi
\frac\partial{\partial\xi}\left(\xi\frac\partial{\partial\xi}\right).
\end{equation}

In this section we will consider the Casimir energy due to parallel
semitransparent plates.  Such a plate is described by a $\delta$-function
potential, $V_i=\lambda_i\delta(\xi-\xi_i)$,
if the plate is perpendicular to the $\xi$ direction.  Note that in
the limit $\lambda\to\infty$ such a potential imposes a Dirichlet
boundary condition on a scalar field.
For two semitransparent
plates, at $\xi_1$ and $\xi_2$, respectively, 
the Green's function can be written as
\begin{equation}
G(x,x')=\int\frac{d\omega}{2\pi}\frac{d^2 k}{(2\pi)^2}e^{-i\omega(\tau-\tau')}
e^{i\mathbf{k\cdot(r-r')_\perp}}g(\xi,\xi'),
\end{equation}
where the reduced Green's function satisfies
\begin{equation}
\left[-\frac{\omega^2}{\xi^2}-\frac1\xi\frac\partial{\partial\xi}\left(\xi\frac
\partial{\partial\xi}\right)+k^2+\lambda_1\delta(\xi-\xi_1)
+\lambda_2\delta(\xi-\xi_2)\right]g=
\frac1\xi\delta(\xi-\xi'),\label{rgfe}
\end{equation}
which we recognize as just the parallel
semitransparent cylinder problem with $m\to \zeta=-i\omega$ and
$\kappa\to k$.

\subsection{Energy-Momentum Tensor}
The canonical energy-momentum for a scalar field is given by
\begin{equation}
T_{\mu\nu}=\partial_\mu\phi\partial_\nu\phi
+g_{\mu\nu}\frac1{\sqrt{-g}}\mathcal{L},
\end{equation}
where the Lagrange density includes the $\delta$-function potentials.  Using
the equations of motion, we find for the energy density,
\begin{equation}
T_{00}=\frac12\left(\frac{\partial\phi}{\partial\tau}\right)^2-\frac12\phi
\frac{\partial^2}{\partial\tau^2}\phi+\frac\xi 2\frac\partial{\partial\xi}
\left(\phi\xi\frac\partial{\partial\xi}\phi\right)
+\frac{\xi^2}2\bnabla_\perp
\cdot(\phi\bnabla_\perp\phi).\end{equation}
Here
$\langle T_{\mu\nu}\rangle$ follows from
$\langle \phi(x)\phi(y)\rangle=\frac1iG(x,y)$.

The force density is given by
\begin{equation}
f_\lambda=-\frac1{\sqrt{-g}}\partial_\nu(\sqrt{-g}T^\nu{}_\lambda)
+\frac12T^{\mu\nu}\partial_\lambda g_{\mu\nu},
\end{equation}
so the gravitational force/area on the system is 
\begin{equation}
\mathcal{F}=\int d\xi\, \xi\, f_\xi=-\int\frac{d\xi}{\xi^2}T_{00}
=\int d\xi \,\xi\int\frac{d\hat\zeta \, d^2k}{(2\pi)^3}\hat\zeta^2 
g(\xi,\xi)\quad(\zeta=\hat\zeta\xi);\label{gf}
\end{equation}
$g$ is  given in terms of modified Bessel
functions.

\subsection{Flat-space Limit for 2 Parallel Plates}
In the weak acceleration limit, $\xi\to\infty$, 
the Green's function corresponding to two parallel plates,
following from (\ref{rgfe}),
reduces to exactly the expected result, for example, for
$\xi_1<\xi,\xi'<\xi_2$
\begin{equation}
\xi_0g(\xi,\xi')\to\frac1{2\kappa}e^{-\kappa|\xi-\xi'|}
+\frac1{2\kappa\tilde\Delta}\bigg[\frac{\lambda_1\lambda_2}
{4\kappa^2}2\cosh\kappa
(\xi-\xi')-\frac{\lambda_1}{2\kappa}\left(1+\frac{\lambda_2}{2\kappa}\right)
e^{-\kappa(\xi+\xi'-2\xi_2)}
-\frac{\lambda_2}{2\kappa}\left(1+\frac{\lambda_1}{2\kappa}\right)
e^{\kappa(\xi+\xi'-2\xi_1)}\bigg],
\end{equation}
where $\kappa^2=k^2+\hat\zeta^2$ and
\begin{equation}
\tilde\Delta=\left(1+\frac{\lambda_1}{2\kappa}\right)
\left(1+\frac{\lambda_2}{2\kappa}\right)e^{2\kappa a}
-\frac{\lambda_1\lambda_2}{4\kappa^2},\quad a=\xi_2-\xi_1.
\end{equation}

\subsection{Explicit Force on 2-plate Apparatus}
If we now use (\ref{gf}), and take the $\xi\to\infty$ limit, we
find the following form for the gravitational force on the 
vacuum energy of the apparatus:
\begin{eqnarray}
\mathcal{F}&=&\frac1{96\pi^2 a^3}\int_0^\infty dy\,y^3
\frac{1+\frac1{y+\lambda_1a}
+\frac1{y+\lambda_2a}}{\left(\frac{y}{\lambda_1a}+1\right)
\left(\frac{y}{\lambda_2a}+1\right)e^y-1}
-\frac1{96\pi^2 a^3}\int_0^\infty dy\,y^2\left[\frac1{\frac{y}{\lambda_1a}
+1}+\frac1{\frac{y}{\lambda_2a}+1}\right]\nonumber\\
&=&-(\mathcal{E}_c+\mathcal{E}_{d1}+\mathcal{E}_{d2}),\end{eqnarray}
which is just the negative of the Casimir energy of the two semitransparent
plates, including divergent parts $\mathcal{E}_{di}$ associated with each
plate.
The divergent terms simply renormalize the bare mass/area $m_i$ of each plate:
\begin{equation}
E_{\rm total}=m_1+m_2+\mathcal{E}_{d1}+\mathcal{E}_{d2}+\mathcal{E}_c
=\mathcal{M}_1+\mathcal{M}_2+\mathcal{E}_c,
\end{equation}
and thus the gravitational force on the entire apparatus obeys the
equivalence principle \cite{Milton:2007ar}
\begin{equation}
g\mathcal{F}_{\rm total}=-g(\mathcal{M}_1+\mathcal{M}_2+\mathcal{E}_c).
\end{equation}

\section{CENTRIPETAL ACCELERATION}
 A Casimir apparatus undergoing centripetal acceleration ($\omega r\ll1$)
also can be shown \cite{Shajesh:2007sc} 
to experience the expected acceleration, in terms of 
$\mathbf{r}_{\rm CM}$, the center of energy of the entire
renormalized system,
\begin{equation}
\mathbf{F}=-\omega^2\int d^3x\,\mathbf{r}\,t^{00}(\mathbf{r})
=-\omega^2\mathbf{r}_{CM}(m_1+m_2+E_{d1}+E_{d2}+E_c).\end{equation}

\section{CONCLUSIONS}
We have demonstrated that
Casimir energy gravitates just like any other form of energy,
$\mathcal{F}=-g\mathcal{E}_c$.
This result, obtained by a variational calculation,
 is independent of the orientation of the Casimir apparatus
relative to the gravitational field. 
Although there was a certain period of confusion \cite{calloni},
there now seems to be complete agreement on this result \cite{Bimonte:2008zv}.
Although gravitational energies have a certain ill-defined character,
being gauge- or coordinate-variant, this result is obtained for a Fermi
observer, the relativistic generalization of an inertial observer.
This conclusion is supported by an explicit calculation in Rindler
coordinates, describing a uniformly accelerated observer.  This demonstrates,
quite generally, that the total Casimir energy, including the divergent
parts, which renormalize the masses of the plates,
possesses the gravitational mass demanded by the equivalence principle.
(For earlier work on the finite part only, see Ref.~\cite{Saharian:2003fd}.)
The inertial properties of Casimir energies
further are confirmed by considering centripetal acceleration.
We hope to consider the acceleration of a Boyer sphere in the near future.

\begin{acknowledgments}
We thank the US National Science Foundation, grant numbers PHY-0554926
and PHY-0554849,
and the US Department of Energy, grant number DE-FG-04ER41305, 
for partially funding this research.
\end{acknowledgments}


\end{document}